\documentclass[conference]{IEEEtran}

\makeatletter
\def\bstctlcite{\@ifnextchar[{\@bstctlcite}{\@bstctlcite[@auxout]}}
\def\@bstctlcite[#1]#2{%
 \@bsphack
 \@for\@citeb:=#2\do{%
 \edef\@citeb{\expandafter\@firstofone\@citeb}%
 \if@filesw\immediate\write\csname #1\endcsname{\string\citation{\@citeb}}\fi}%
 \@esphack}
\makeatother

\IEEEoverridecommandlockouts
\usepackage{cite}
\usepackage{amsmath,amssymb,amsfonts}
\usepackage[cmintegrals]{newtxmath}
\usepackage{algorithm}
\usepackage{algorithmic}
\usepackage{array}
\usepackage[caption=false,font=footnotesize]{subfig}
\usepackage{stfloats}
\usepackage{url}
\usepackage{graphicx}
\usepackage{textcomp}
\usepackage{xcolor}
\usepackage{dane_nmath}
\usepackage{tikz}
\usepackage{ellipsis}
\usetikzlibrary{calc}
\usetikzlibrary{decorations.pathreplacing,decorations.markings,shapes.geometric}
\usetikzlibrary{chains}

\tikzstyle{block} = [draw, rectangle, 
    minimum height=4em, minimum width=4em]
\tikzstyle{input} = [coordinate]
\tikzstyle{output} = [coordinate]
\tikzstyle{pinstyle} = [pin edge={to-,thin,black}]

\usetikzlibrary{positioning}

\tikzset{radiation/.style={{decorate,decoration={expanding waves,angle=90,segment length=5pt}}}}
\tikzset{font={\fontsize{9.5pt}{12}\selectfont}}

\usetikzlibrary{spy}
\usepackage{pgfplots}
\usepackage{wrapfig}
\usetikzlibrary{arrows,shapes}
\usetikzlibrary{positioning,shapes.callouts}
\usepgfplotslibrary{groupplots,dateplot}
\usetikzlibrary{patterns,shapes.arrows}
\pgfplotsset{compat=newest}

\def\BibTeX{{\rm B\kern-.05em{\sc i\kern-.025em b}\kern-.08em
    T\kern-.1667em\lower.7ex\hbox{E}\kern-.125emX}}

\newcommand{\wout}[1]{\textcolor{black}{#1}}
\newcommand{\vari}[1]{\textcolor{black}{#1}}

\newcommand{\wo}[1]{\textcolor{black}{#1}}

\newcommand{\csubfloat}[2][]{%
  \makebox[0pt]{\subfloat[#1]{#2}}%
}

\newcommand{\varizz}[1]{\textcolor{black}{#1}}

\newcommand\copyrighttext{%
	\footnotesize \textcopyright 2021 IEEE 22nd {IEEE} International Workshop on Signal Processing Advances in Wireless Communications, {SPAWC} 2021, Lucca, Italy. Personal use of this material is permitted.  Permission from IEEE must be obtained for all other uses, in any current or future media, including reprinting/republishing this material for advertising or promotional purposes, creating new collective works, for resale or redistribution to servers or lists, or reuse of any copyrighted component of this work in other works.}
\newcommand\copyrightnotice{%
	\begin{tikzpicture}[remember picture,overlay]
	\node[anchor=south,yshift=10pt] at (current page.south) {\fbox{\parbox{\dimexpr\textwidth-\fboxsep-\fboxrule\relax}{\copyrighttext}}};
	\end{tikzpicture}%
}

\begin{document}
\bstctlcite{IEEEexample:BSTcontrol}
\title{Learning the CSI Denoising and Feedback\\Without Supervision
\thanks{This research was supported by an unrestricted gift from Futurewei Technologies, Inc., Huawei R\&D USA.}
}
\newcommand{\jbig}{$\mathcal{J}$}
\author{\IEEEauthorblockN{Valentina Rizzello \wout{and} Wolfgang Utschick \\}
\IEEEauthorblockA{\textit{Department of Electrical and Computer Engineering} \\
\textit{Technical University of Munich} \\
\small{\texttt{\{valentina.rizzello, utschick\}@tum.de}}}
}

\maketitle
\copyrightnotice
\begin{abstract}
In this work, we develop a joint denoising and feedback strategy for channel state information in frequency division duplex systems. In \wout{such systems, the biggest challenge is the overhead incurred when the mobile terminal has to send the downlink channel state information or corresponding partial information to the base station, where the complete estimates can subsequently be restored. To this end, we propose a novel learning-based framework for denoising and compression of channel estimates.
Unlike existing studies, 
we extend a recently proposed approach 
and \wout{show that based solely on noisy uplink data available at the base station,} it is possible to learn an autoencoder neural network that generalizes to downlink data. Subsequently, half of the autoencoder can be offloaded to the \wo{mobile terminals} to generate channel feedback there as efficiently as possible, without any training effort at the \wo{terminals} or corresponding transfer of training data. Numerical simulations demonstrate the excellent performance of the proposed method.}
\end{abstract}


\section{Introduction}
Massive multiple-input multiple-output (MIMO) is certainly the most noticeable technology to increase the throughput and guarantee reliability in modern and future wireless communication systems~\cite{marzetta}. With \wout{the deployment of} large-scale antenna arrays, space diversity induces a remarkable improvement in the spectral efficiency and makes possible to serve multiple users at the same time. However, to benefit from all the \wout{prospective} advantages of massive MIMO, the high dimensional channel frequency response must be accurate and promptly acquired at the \wout{base station} (BS). 
Therefore, the strong reciprocity between the \wout{corresponding uplink (UL) and downlink (DL) channels} makes time division duplex (TDD) networks \wout{one of the most prominent solution candidates} under these strict constraints~ \cite{sanguinetti2019massive}. In contrast, in \wout{frequency division duplex} (FDD) systems, the absence of reciprocity between \wout{the} UL and DL channel responses and consequently the huge overhead for reporting the \wout{channel state information (CSI)} from the \wout{mobile terminal} (MT) to the BS represent\wout{s} the major limitation for an effective deployment of massive MIMO communications.
Although the TDD operation mode is the most commonly adopted, it has been shown that FDD massive MIMO would handle the low latency requirements imposed by the standardization potentially much better than TDD solutions~\cite{BjornsonLM16}.
Hence, this premise has motivated and encouraged several studies that aim to reduce or eliminate the DL CSI acquisition overhead. \wout{In addition to some well-known examples based on a particular model and sparsity assumptions that show how to extrapolate DL covariance from UL covariance~\cite{8542957}, there are a variety of data-driven approaches that address the challenge of recovering instantaneous DL CSI in FDD systems at the BS.} \wout{Among these, to eliminate the need for feedback, several machine learning approaches have been proposed based on supervised learning of direct extrapolation of CSI across the frequency gap, based on pairs of UL--DL training data}~\cite{ArDoCaYaHoBr19,alk2019deep,8764345,han2020deep,safari,me}. \\
\indent\wout{A very innovative solution is represented by the concept of autoencoder neural networks which are trained in order to learn a low rate feedback from the MT to the BS}~\cite{csinet,8638509,9090892,8972904,9279228,9347820}. In this setup, the DL CSI is encoded at the MT into a codeword, which is then fed back to the BS \wout{and decoded there, implying a distributed implementation of the parts of the autoencoder at the MT and BS.} \\
\indent\wo{In this work, following this general approach, we propose a novel method which is {again} based on {the autoencoding concept. However, motivated by the results in~\cite{utschick2021learning}, the unsupervised training of the autoencoder is conducted at the BS soley based on noisy UL training data, thus avoiding the issue that collecting DL data at the BS to enable the training otherwise would require an immense effort with respect to the overall network traffic.} 
{By the corresponding result from \cite{utschick2021learning}, we mean the equivalence of UL and DL CSI discovered therein with respect to their probability distributions.}
Thus, the core idea of our scheme is that the neural network encoder trained on UL data at the BS can be applied to DL data without any further adaptation, from any mobile device to which the encoder is offloaded. Training on the MT is no longer necessary at all, making it possible to quickly update the encoder on the MT at any time and place, e.g., when moving from one cell to another or for different locations in the cell.} 
\varizz{Compared to our approach, training at the MT with DL data has some disadvantages, e.g.: \textit{i)} the MT could spend only a short amount of time inside a cell and could not collect enough samples for training, \textit{ii)} if multiple MTs stayed in the same cell long enough to perform the training of different autoencoders, lots of computational power would be wasted since only one decoder would be deployed at the BS, \textit{iii)} there would be a high risk of overfitting since it's unlikely that a MT visits all the locations in a cell because of the systematic behaviour of the users.}
\wout{Based on the presented simulation results, we are eventually able to demonstrate the excellent performance of the proposed technique.}

\section{System Architecture}
In the following, we indicate with $\ulnoisy$ and $\dlnoisy \in \mathbb{C}^{N_{\text{a}}\times N_{\text{c}}}$ the noisy UL and DL CSI matrices \wout{of the transmission channel between the BS and the single antenna MT}, where $N_{\text{a}}$ and $N_{\text{c}}$ denote the
number of antennas at the BS and the number of subcarriers, respectively. In addition, we can express $\ulnoisy$ as
\begin{equation} \ulnoisy = \ulbig + \vN,
\end{equation}
\vari{where $\ulbig$ and $\vN \in \mathbb{C}^{N_{\text{a}}\times N_{\text{c}}}$ represents the true UL CSI matrix and the additive white Gaussian noise, respectively. 
Analogous expressions can be derived for $\dlnoisy$. }
\wout{Note that throughout} this work we assume that the true data for both UL and DL, namely $\ulbig$ and $\dlbig$, are inaccessible and only a noisy version of them is available.
\begin{figure}[t]
\centering
  \hspace{-1cm}
  \csubfloat[\wout{Training of the autoencoder at the BS.}]{\label{fig: aeUL}
    \definecolor{mycolor}{HTML}{7B241C}

\tikzset{radiation/.style={{decorate,decoration={expanding waves,angle=90,segment length=5pt}}}}
\tikzset{radiation1/.style={{decorate,decoration={expanding waves,angle=90,segment length=19pt}}}}
\def\mobilestation{
\begin{tikzpicture}
 \draw[thick] 
 (1.3, -1.3) -- (1.3, -0.3)
 (1.9, -1.3) -- (1.9, -0.3)
 (1.9, -0.3) -- (1.3, -0.3)
 (1.3, -1.3) -- (1.9, -1.3);
  \draw (1.3, -0.4) -- (1.0, -0.4) (1.0, -0.4)--(1.0, -0.1) (1.0, -0.1)--(0.8, 0.15) (1.0, -0.1)--(1.2, 0.15) (1.2, 0.15)--(0.8, 0.15);
\end{tikzpicture}
}

\def\basestation{
\begin{tikzpicture}
 \draw[thick] 
    (-0.4, 0.2) -- (-0.4, -0.2)
    (0.4, 0.2) -- (0.4, -0.2)
    (0, 0.4) -- (0, 0)
    (-0.4, 0) -- (0.4, 0)
    (-0.9, -2.7) -- (0, 0) coordinate[pos=0.05] (l3)
    coordinate[pos=0.35] (l4)
        coordinate[pos=0.55] (l2) coordinate[pos=0.65] (l1)
    (0.9, -2.7) -- (0, 0) coordinate[pos=0.05] (r3)
    coordinate[pos=0.35] (r4)
        coordinate[pos=0.55] (r2) coordinate[pos=0.65] (r1);
  \draw (l1) -- (r2) (l2)--(r1) (l3)--(r4) (r3)--(l4);
  \draw[radiation,decoration={angle=45}] (0, 0.4) -- ++ (90:0.7cm);
  \draw (0.7, -0.4) -- (1.0, -0.4) (1.0, -0.4)--(1.0, -0.1) (1.0, -0.1)--(0.8, 0.15) (1.0, -0.1)--(1.2, 0.15) (1.2, 0.15)--(0.8, 0.15);
  \draw (0.7, 0.3) -- (1.0, 0.3) (1.0, 0.3)--(1.0, 0.6) (1.0, 0.6)--(0.8, 0.85) (1.0, 0.6)--(1.2, 0.85) (1.2, 0.85)--(0.8, 0.85);
  \draw[thick, dotted] (1.0, -0.6) -- (1.0, -0.9);
  \draw (0.7, -1.6) -- (1.0, -1.6) (1.0, -1.6)--(1.0, -1.3) (1.0, -1.3)--(0.8, -1.05) (1.0, -1.3)--(1.2, -1.05) (1.2, -1.05)--(0.8, -1.05);
\end{tikzpicture}
}

\definecolor{my_color}{HTML}{4D5656}
\definecolor{fill_c}{HTML}{D6EAF8}
\def\mobilestationnew{\begin{tikzpicture}[line width=4pt, color=my_color]
\draw[rounded corners=0.5cm, fill=fill_c](0,0) rectangle (6.73,13.84);
\draw[rounded corners=0.07cm, fill=white,line width=3.5pt](2.765,13.045) rectangle (3.965,13.235);
\draw[fill=white,](2.4,13.13) circle (0.15);
\draw[fill=white, line width=4.2pt](3.365,0.75) circle (0.5);
\draw[rounded corners=0.04cm] (2.366,0.466);
\draw[fill=white, line width=4.2pt](0.5,1.5) rectangle (6.23,12.44);
 \draw[radiation1,decoration={angle=45}, line width=4.8pt] (6.9, 14) -- ++ (50:2.5cm);
\end{tikzpicture}
}

\definecolor{my_color2}{HTML}{4D5656}
\definecolor{fill_c2}{HTML}{F5EEF8}

\def\basestationnew{\begin{tikzpicture}[line width=2pt, black]
    \draw[rounded corners=0.07cm, line width=2pt, fill=babypink!30](-0.6, 0) rectangle (0.6, 1.2);
    \draw[thick, line width=2pt] (-0.9, -2.7) -- (0, 0) coordinate[pos=0.05] (l3)
    coordinate[pos=0.35] (l4)
        coordinate[pos=0.55] (l2) coordinate[pos=0.65] (l1)
    (0.9, -2.7) -- (0, 0) coordinate[pos=0.05] (r3)
    coordinate[pos=0.35] (r4)
        coordinate[pos=0.55] (r2) coordinate[pos=0.65] (r1);
  \draw (l1) -- (r2) (l2)--(r1) (l3)--(r4) (r3)--(l4);
  \draw[fill=babypink, line width=1.3pt](-0.3, 0.3) circle (0.1);
  \draw[fill=babypink, line width=1.3pt](0.3, 0.9) circle (0.1);
  \draw[fill=babypink, line width=1.3pt](-0.3, 0.9) circle (0.1);
    \draw[fill=babypink, line width=1.3pt](0.3, 0.6) circle (0.1);
  \draw[fill=babypink, line width=1.3pt](-0.3, 0.6) circle (0.1);
  \draw[fill=babypink, line width=1.3pt](0.3, 0.3) circle (0.1);
   \draw[fill=babypink, line width=1.3pt](0, 0.9) circle (0.1);
    \draw[fill=babypink, line width=1.3pt](0, 0.6) circle (0.1);
    \draw[fill=babypink, line width=1.3pt](0, 0.3) circle (0.1);
  \draw[radiation,decoration={angle=45}, line width=2pt] (0.7, 0.6) -- ++ (-6:0.7cm);
\end{tikzpicture}
}

\tikzset{>=latex}

\definecolor{babyblue}{rgb}{0.54, 0.81, 0.94}
\definecolor{babypink}{rgb}{0.96, 0.76, 0.76}
\begin{tikzpicture}[auto, node distance=2.5cm, line width=0.8pt, scale=0.6, every node/.style={scale=0.9}]
    \node [input, name=input] at (0, 0) {};
    \node[fill=babypink,draw, trapezium, rotate=90, minimum height=0.9cm, trapezium stretches body, left of=input] (dec) {\rotatebox{-90}{Dec}};
        \node[fill=babypink,draw, trapezium, rotate=-90, minimum height=0.9cm, trapezium stretches body, right=-1.6cm of dec] (enc) {\rotatebox{90}{Enc}};
    \node(v1)[left of=enc]{$\tilde{\vect{H}}_{\text{UL}}$@BS};
    \node(v2)[right of=dec]{$\def\arraystretch{0.7} \begin{array}{c} \ulbighat \approxeq \\ \ulbig\text{@BS}\end{array}$};
    \node[below left =-0.05cm and -0.2cm of v2, scale=0.35]{
    \basestationnew
    };
    \node[below right =0.4cm and 1cm of v1]{
    $ \vf_{\boldsymbol{\theta}}(\cdot)$
    };
    \node[below right =0.4cm and 2.6cm of v1]{    
    $ \vg_{\boldsymbol{\phi}}(\cdot) $
    };
    
    
    \draw[->, >=stealth] (v1)--(enc);
    \draw[->, >=stealth] (dec)--(v2);
    \draw[->, >=stealth] (enc)--(dec);
\end{tikzpicture}}\\
\hspace{-1.5cm}
  \csubfloat[\wout{Codeword generation at the MT.}]{\label{fig: aeDL}
\tikzset{radiation/.style={{decorate,decoration={expanding waves,angle=90,segment length=5pt}}}}
\tikzset{radiation1/.style={{decorate,decoration={expanding waves,angle=90,segment length=19pt}}}}
\definecolor{babyblue}{rgb}{0.54, 0.81, 0.94}
\definecolor{babypink}{rgb}{0.96, 0.76, 0.76}

\def\mobilestation{
\begin{tikzpicture}
 \draw[thick] 
 (1.3, -1.3) -- (1.3, -0.3)
 (1.9, -1.3) -- (1.9, -0.3)
 (1.9, -0.3) -- (1.3, -0.3)
 (1.3, -1.3) -- (1.9, -1.3);
  \draw (1.3, -0.4) -- (1.0, -0.4) (1.0, -0.4)--(1.0, -0.1) (1.0, -0.1)--(0.8, 0.15) (1.0, -0.1)--(1.2, 0.15) (1.2, 0.15)--(0.8, 0.15);
\end{tikzpicture}
}

\def\basestation{
\begin{tikzpicture}
 \draw[thick] 
    (-0.4, 0.2) -- (-0.4, -0.2)
    (0.4, 0.2) -- (0.4, -0.2)
    (0, 0.4) -- (0, 0)
    (-0.4, 0) -- (0.4, 0)
    (-0.9, -2.7) -- (0, 0) coordinate[pos=0.05] (l3)
    coordinate[pos=0.35] (l4)
        coordinate[pos=0.55] (l2) coordinate[pos=0.65] (l1)
    (0.9, -2.7) -- (0, 0) coordinate[pos=0.05] (r3)
    coordinate[pos=0.35] (r4)
        coordinate[pos=0.55] (r2) coordinate[pos=0.65] (r1);
  \draw (l1) -- (r2) (l2)--(r1) (l3)--(r4) (r3)--(l4);
  \draw[radiation,decoration={angle=45}] (0, 0.4) -- ++ (90:0.7cm);
  \draw (0.7, -0.4) -- (1.0, -0.4) (1.0, -0.4)--(1.0, -0.1) (1.0, -0.1)--(0.8, 0.15) (1.0, -0.1)--(1.2, 0.15) (1.2, 0.15)--(0.8, 0.15);
  \draw (0.7, 0.3) -- (1.0, 0.3) (1.0, 0.3)--(1.0, 0.6) (1.0, 0.6)--(0.8, 0.85) (1.0, 0.6)--(1.2, 0.85) (1.2, 0.85)--(0.8, 0.85);
  \draw[thick, dotted] (1.0, -0.6) -- (1.0, -0.9);
  \draw (0.7, -1.6) -- (1.0, -1.6) (1.0, -1.6)--(1.0, -1.3) (1.0, -1.3)--(0.8, -1.05) (1.0, -1.3)--(1.2, -1.05) (1.2, -1.05)--(0.8, -1.05);
\end{tikzpicture}
}

\definecolor{my_color}{HTML}{4D5656}
\definecolor{fill_c}{HTML}{D6EAF8}

\def\mobilestationnew{\begin{tikzpicture}[line width=6pt, color=black]
\draw[rounded corners=0.5cm,line width=7.5pt, fill=babyblue!30](0,0) rectangle (6.73,13.84);
\draw[rounded corners=0.07cm, fill=white,line width=4pt](2.765,13.045) rectangle (3.965,13.235);
\draw[fill=babyblue,line width=4pt](2.4,13.13) circle (0.15);
\draw[fill=babyblue, line width=5.5pt](3.365,0.75) circle (0.5);
\draw[rounded corners=0.04cm] (2.366,0.466);
\draw[fill=babyblue, line width=7.5pt](0.5,1.5) rectangle (6.23,12.44);
 \draw[radiation1,decoration={angle=45}, line width=8pt] (6.9, 14) -- ++ (50:2.5cm);
\end{tikzpicture}
}

\definecolor{my_color2}{HTML}{873600}
\definecolor{fill_c2}{HTML}{F5EEF8}

\def\basestationnew{\begin{tikzpicture}[line width=2pt, black]
    \draw[rounded corners=0.07cm, line width=2pt, fill=babypink!30](-0.6, 0) rectangle (0.6, 1.2);
    \draw[thick, line width=2pt] (-0.9, -2.7) -- (0, 0) coordinate[pos=0.05] (l3)
    coordinate[pos=0.35] (l4)
        coordinate[pos=0.55] (l2) coordinate[pos=0.65] (l1)
    (0.9, -2.7) -- (0, 0) coordinate[pos=0.05] (r3)
    coordinate[pos=0.35] (r4)
        coordinate[pos=0.55] (r2) coordinate[pos=0.65] (r1);
  \draw (l1) -- (r2) (l2)--(r1) (l3)--(r4) (r3)--(l4);
  \draw[fill=babypink, line width=1.3pt](-0.3, 0.3) circle (0.1);
  \draw[fill=babypink, line width=1.3pt](0.3, 0.9) circle (0.1);
  \draw[fill=babypink, line width=1.3pt](-0.3, 0.9) circle (0.1);
    \draw[fill=babypink, line width=1.3pt](0.3, 0.6) circle (0.1);
  \draw[fill=babypink, line width=1.3pt](-0.3, 0.6) circle (0.1);
  \draw[fill=babypink, line width=1.3pt](0.3, 0.3) circle (0.1);
   \draw[fill=babypink, line width=1.3pt](0, 0.9) circle (0.1);
    \draw[fill=babypink, line width=1.3pt](0, 0.6) circle (0.1);
    \draw[fill=babypink, line width=1.3pt](0, 0.3) circle (0.1);
  \draw[radiation,decoration={angle=45}, line width=2pt] (0.7, 0.6) -- ++ (-6:0.7cm);
\end{tikzpicture}
}

\tikzset{>=latex}
\tikzset{
  zigzag/.style={to path={ -- ($(\tikztostart)!.55!-7:(\tikztotarget)$) -- ($(\tikztostart)!.45!7:(\tikztotarget)$) -- (\tikztotarget) \tikztonodes}}
}


\begin{tikzpicture}[auto, node distance=2.5cm, line width=0.8pt, scale=0.6, every node/.style={scale=0.9}]
    \node [input, name=input] at (0, 0) {};
    \node[fill=babypink,draw, trapezium, rotate=90, minimum height=0.9cm, trapezium stretches body, left of=input] (dec) {\rotatebox{-90}{Dec}};
        \node[fill=babyblue,draw, trapezium, rotate=-90, minimum height=0.9cm, trapezium stretches body, right=-2.6cm of dec] (enc) {\rotatebox{90}{Enc}};
    \node(v1)[left of= enc]{$\tilde{\vect{H}}_{\text{DL}}$@MT};
    \node(v2)[right of=dec]{$\def\arraystretch{0.7} \begin{array}{c} \dlbighat \approxeq \\ \dlbig\text{@BS}\end{array}$};
    \node[below right =0.26cm and 0cm of v1, scale=0.08]{
    \mobilestationnew
    };
    \node[below left =-0.05cm and -0.2cm of v2, scale=0.35]{
    \basestationnew
    };
    \node[above right =0.4cm and 0.5cm of v1]{
    $\def\arraystretch{0.7} \begin{array}{c} \text{offloaded} \\ \text{from BS:} \\  \vf_{\boldsymbol{\theta}}(\cdot) \end{array} $
    };
    
    
    \draw[->, >=stealth] (v1)--(enc);
    \draw[->, >=stealth] (dec)--(v2);
    \draw[->, >=stealth, color=black] (enc) to[zigzag] (dec);
    \node[below right=-1.6cm and 0.9cm of enc, scale=1]{
    $ \def\arraystretch{0.7}\begin{array}{c} \text{CSI} \\ \text{feedback}\end{array} $
    };
\end{tikzpicture}

}
  \caption{\wout{Training of the autoencoder based on UL CSI at the BS, generation of the codeword by the offloaded encoder at the MT, transmission over the radio channel, and subsequent reconstruction of the DL CSI at the BS.}}
  \label{fig: sole_ae}
  \vspace*{-4mm}
\end{figure}
The proposed method consists of two \wout{phases, which are} illustrated in Fig.~\ref{fig: sole_ae}. \wout{First, an autoencoder $\vg_{\boldsymbol{\phi}} (\vf_{\boldsymbol{\theta}} (\cdot)) $ is trained at the BS based solely on noisy UL data $\ulnoisy$, which is supposed to be collected during the standard UL operation of the BS in advance.} The $\vf_{\boldsymbol{\theta}}$ denotes the encoder with parameters $\boldsymbol{\theta}$ and $\vg_{\boldsymbol{\phi}}$ denotes the decoder with parameters $\boldsymbol{\phi}$, see Fig.~\ref{fig: aeUL}. \wout{It is well-known that autoencoders implicitly introduce regularization for the reconstruction of the input signal, cf.~\cite{10.1093/imaiai/iaaa011} for an introduction to the fundamentals behind denoising with deep neural networks.}
In essence, \wout{an} autoencoder can be trained with the noisy data $\ulnoisy$ in an unsupervised fashion 
to obtain an estimate $\ulbighat$ which will be approximately equal to the unknown $\ulbig$. 
\wout{It should be noted that for the proposed method, there are no special requirements for the acquisition of the UL training data, except for the property that they come from the same propagation scenario as the subsequent DL data to which the encoder \wo{will be applied at the MTs}.}
\wout{Subsequently, half of the autoencoder, namely the encoding part $\vf_{\boldsymbol{\theta}}(\cdot)$, is offloaded to the MT based on a respective network protocol, which is due to space restriction not further considered here.} 

\wout{In the second phase,} similarly to what has been proposed in~\cite{utschick2021learning}, we reuse the UL-trained autoencoder neural network for the recovery \wout{of the complete DL CSI.}
In particular, \wo{each} MT takes the noisy DL CSI estimate $\dlnoisy$ and feeds it into 
the \wout{offloaded} UL-trained encoder to obtain the latent vector or codeword $\vz_{\text{DL}}$. Then, the codeword is fed back to the BS which recovers $\dlbighat \approxeq \dlbig$ with the second half of the autoencoder, namely the UL-trained decoder.

\section{Dataset Description}
\label{sec: quad}
Our study is based on a single urban microcell (UMi) with $150$~\wout{meters} radius, which has been simulated with the $\Matlab$ based software QuaDRiGa version 2.2~\cite{quad, quad2}.
Specifically, we consider non-line-of-sight (NLoS) channels, with $L=58$  multi-path components (MPCs), \wout{which means a rich scattering propagation environment.} The BS is placed at a height of $10$ meters and is equipped with a
uniform planar array (UPA) with $N_{\text{a}} =  8\times 8$ ``3GPP-3d'' antennas, while the
users have a single omni-directional antenna each. In addition, the BS antennas
are tilted by $6$ degrees towards the ground to point in the direction of the
users.
The UL center frequency is $2.5$~GHz while the DL center frequencies are
$2.62$~GHz, and $2.98$~GHz, which correspond to a FDD gap of $120$~MHz and $480$~MHz, respectively. For each frequency, we consider a bandwidth of approximately $8$~MHz divided over $ N_{\text{c}} =  160 $ subcarriers.
The cell has been sampled at \wout{$60\times 10^3$} \wo{different locations of MT} and for each sample the channels at the
predefined frequencies are collected.
Therefore, the dataset is split into three groups of \wout{$48\times 10^3$, $6\times 10^3$ and $6\times 10^3$} samples,
where each sample consists of the three matrices ${\vH}_{\text{UL}} $, ${\vH}_{\text{DL-120}} $, and ${\vH}_{\text{DL-480}} \in \mathbb{C}^{N_{\text{a}}\times N_{\text{c}}} $.
\wout{Note again that although the training of the autoencoder at the BS is based solely on the UL CSI, it still covers the distribution of the unseen DL CSI as well, since the UL and DL data ultimately follow the same propagation scenario, cf. ~\cite{utschick2021learning}. With respect to testing, only the test set of the two DL CSI datasets (DL@120, 480) will be used.} Additionally, and likewise~\cite{utschick2021learning} the channels are normalized with respect to their path-gain.

\section{Autoencoder}
\label{sec: ae}
An autoencoder is a neural network that is trained in an unsupervised fashion to reconstruct its input. It has been introduced in~\cite{aebook} and its purpose is to find a compact representation of the data.
The autoencoder consists of two parts: an encoder function $\vf_{\boldsymbol{\theta}}$ with hyperparameters $\boldsymbol{\theta}$ and a decoder function $\vg_{\boldsymbol{\phi}}$ with hyperparameters $\boldsymbol{\phi}$. 
The encoder projects a $d$-dimensional input vector $\vx$ into a \wout{typically lower} dimensional latent \wout{space representation $\vz \in \mathbb{C}^{d_z} $ with $d_z \ll d$,} whereas the decoder reconstructs the original input from $\vz$, \wout{i.e.,
\begin{equation}
    \boldsymbol{x} \stackrel{\vf_{\boldsymbol{\theta}}}{\longrightarrow} 
    \boldsymbol{z}
    \stackrel{\vg_{\boldsymbol{\phi}}}{\longrightarrow} \hat{\vx} \approxeq \vx.
\end{equation}}
Note that the bottleneck or \wout{hourglass structure of the architecture is a key element of the autoencoding concept, as it forces the network to learn only the important features that allow reconstruction with the decoder, cf.}~\cite{Goodfellow-et-al-2016} and~\cite{bank2021autoencoders}.
\begin{table}[t]
  \caption{Encoder architecture.}
  \label{tab: cnn_enc}
  \begin{center}
  \begin{tabular}{lcccr}
  \hline
  Layer type& Output shape & \#Parameters $\boldsymbol{\theta} $\\
  \hline
  Input & $64 \times 160 \times 2$ & 0\\
  Conv2D, strides=2 & $32 \times 80 \times 8$ & 152\\
  Batch normalization & $32 \times 80 \times 8$ & 32\\
  ReLU & $32 \times 80 \times 8$ & 0\\
  Conv2D, strides=2 & $16 \times 40 \times 16$ & 1168\\
  Batch normalization & $16 \times 40 \times 16$ & 64\\
  ReLU & $16 \times 40 \times 16$ & 0\\
  Conv2D, strides=2 & $8 \times 20 \times 32$ & 4640\\
  Batch normalization & $8 \times 20 \times 32$ & 128\\
  ReLU & $8 \times 20 \times 32$ & 0\\
  Conv2D, strides=2 & $4 \times 10 \times 64$ & 18496\\
  Batch normalization & $4 \times 10 \times 64$  & 256\\
  ReLU & $4 \times 10 \times 64$ & 0\\
  Conv2D, strides=2 & $2 \times 5 \times 128$ & 73856\\
  Batch normalization & $2 \times 5 \times 128$ & 512 \\
  ReLU & $2 \times 5 \times 128$ & 0\\
  Flatten & $1280$ & 0\\
  Fully-connected & $256$ & 327936\\
  Tanh & $256$ & 0\\
  \hline
\end{tabular}
\vspace*{-4mm}
\end{center}
\end{table}

\begin{table}[t]
  \caption{Decoder architecture.}
  \label{tab: cnn_dec}
  \begin{center}
  \begin{tabular}{lcccr}
  \hline
  Layer type& Output shape & \#Parameters $\boldsymbol{\phi} $\\
  \hline
  Input & $256$ & 0\\
  Fully-connected & $1280$ & 328960\\
  Reshape & $2\times 5 \times 128$ & 0\\
  Conv2D transposed, strides=2 & $4 \times 10 \times 128$ & 147584\\
  Batch normalization & $4 \times 10 \times 128$  & 512\\
  ReLU & $4 \times 10 \times 128$ & 0\\
  Conv2D transposed, strides=2 & $8 \times 20 \times 64$ & 73792\\
  Batch normalization & $8 \times 20 \times 64$ & 256\\
  ReLU & $8 \times 20 \times 64$ & 0\\
  Conv2D transposed, strides=2 & $16 \times 40 \times 32$  & 18464\\
  Batch normalization & $16 \times 40 \times 32$  & 128\\
  ReLU & $16 \times 40 \times 32$ & 0\\
  Conv2D transposed, strides=2 & $32 \times 80 \times 16$  & 4624\\
  Batch normalization & $32 \times 80 \times 16$  & 64\\
  ReLU & $32 \times 80 \times 16$ & 0\\
  Conv2D transposed, strides=2 & $64 \times 160 \times 8$  & 1160\\
  Batch normalization & $64 \times 160 \times 8$  & 32\\
  ReLU & $64 \times 160 \times 8$ & 0\\
  Conv2D transposed & $64 \times 160 \times 2$  & 146\\
  \hline
\end{tabular}
\vspace*{-4mm}
\end{center}
\end{table}
For \wout{the proposed autoencoder in this work,} we use a deep neural network with several convolutional layer\wout{s}. The encoder and decoder architectures are described in Tables~\ref{tab: cnn_enc} and~\ref{tab: cnn_dec}. Firstly, the real and imaginary parts of the original noisy UL matrix $\ulnoisy \in \mathbb{C}^{64 \times 160} $ have been stacked along the third dimension to form a real-valued tensor $\ulnoisy^{\text{real}} \in \mathbb{R}^{64 \times 160\times 2} $, which represents the input of the encoder. By observing the encoder in Table~\ref{tab: cnn_enc}, we can distinguish \wout{five} consecutive blocks, each of them formed by the cascade of a convolutional layer, a batch normalization layer~\cite{SanturkarTIM18}, and the rectified linear unit (ReLU) activation function. A key attribute of this architecture is \wout{to} use strided convolutions~\cite{SpringenbergDBR14} which are meant to progressively extract features and reduce the input dimension down to $1280$ units. After the progressive reduction of the input dimension, a fully connected layer with \wout{$\tanh(\cdot)$ activation functions} completes the encoder and generates the codeword $\vz_{\text{UL}}$, which is a real valued vector with $d_z = 256$ dimensions that leads to a compression factor \wout{of} 
\begin{equation} 
\frac{64\times 160 \times 2}{256} = 80.
\end{equation} 
Note that having a deep architecture with multiples strided convolutional layers before the fully connected layer helps to substantially reduce the total number of trainable parameters which is highly affected by the number of parameters in the fully connected layer. The decoder, which is displayed in Table~\ref{tab: cnn_dec}, \wout{is supposed to map} the codeword back to the original input $\ulnoisy^{\text{real}} $, \wout{thereby benefiting from the regularizing effect (denoising) of the autoencoder concept. Its structure is equal to} the mirrored version of the encoder, where deconvolutions are in place of convolutions and a final transposed convolution with \wout{two} feature maps recovers the original input shape.
\varizz{Despite the large size, this autoencoder architecture has a number of trainable parameters which is smaller compared to autoencoders built with the same principle of CsiNet~\cite{csinet}.}
\begin{figure*}[t!]
   \centering
  \subfloat[][CDFs NMSE.]{\label{fig: cdf_nmse}
    \input{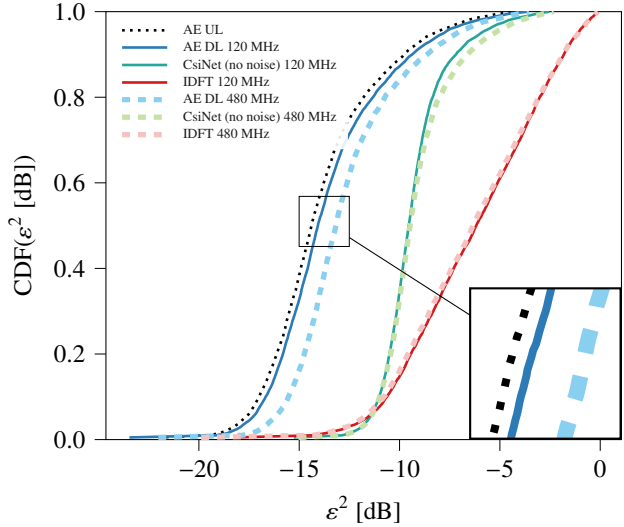}}\hspace{10pt}
  \subfloat[][CDFs Cosine Similarity.]{\label{fig: cdf_rho}\input{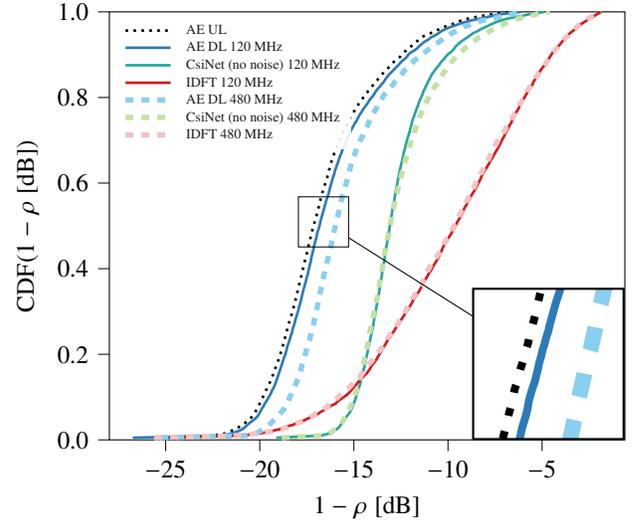}}
  \caption{CDFs performance metrics of different methods for $\text{SNR}=10$~dB.}
  \label{fig: cdfs}
\end{figure*}

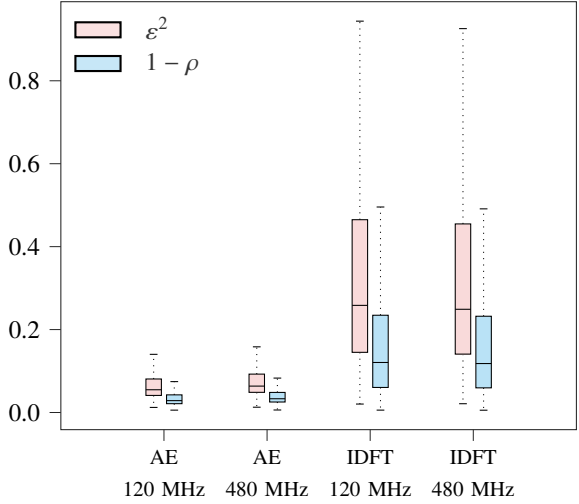
\begin{figure}
    \centering
\begin{tikzpicture}


\definecolor{color0}{HTML}{F9E79F}
\definecolor{color1}{HTML}{85C1E9}

\definecolor{babyblue}{rgb}{0.54, 0.81, 0.94}
\definecolor{babypink}{rgb}{0.96, 0.76, 0.76}

\begin{axis}[
axis line style={white!15!black},
legend cell align={left},
legend style={fill opacity=0.8, draw opacity=1, text opacity=1, at={(0.01,0.99)}, anchor=north west, draw=none},
tick align=outside,
tick pos=left,
x grid style={white!80!black},
xmin=-2, xmax=8,
xtick style={color=white!15!black},
xtick={0,2,4,6},
xticklabels={{\footnotesize{AE}\\\footnotesize{120 MHz}},{\footnotesize{AE}\\\footnotesize{480 MHz}},{\footnotesize{IDFT}\\\footnotesize{120 MHz}},{\footnotesize{IDFT}\\\footnotesize{480 MHz}}},
xticklabel style   = {align=center},
y grid style={white!80!black},
ymin=-0.041371044527176, ymax=0.990711515249261,
ytick style={color=white!15!black},
ytick={-0.2,0,0.2,0.4,0.6,0.8,1},
yticklabels={−0.2,0.0,0.2,0.4,0.6,0.8,1.0}
]

\addlegendimage{area legend, thick,fill = babypink!60}
\addlegendentry{$\varepsilon^2$}

\addlegendimage{area legend, thick,fill = babyblue!60}
\addlegendentry{$1-\rho$}

\addplot [black, dotted, forget plot]
table {%
-0.2 0.0411806143820286
-0.2 0.0122574539855123
};
\addplot [black, dotted, forget plot]
table {%
-0.2 0.0808447673916817
-0.2 0.140283152461052
};
\addplot [black, forget plot]
table {%
-0.275 0.0122574539855123
-0.125 0.0122574539855123
};
\addplot [black, forget plot]
table {%
-0.275 0.140283152461052
-0.125 0.140283152461052
};
\addplot [black, dotted, forget plot]
table {%
1.8 0.0486404057592154
1.8 0.012775681912899
};
\addplot [black, dotted, forget plot]
table {%
1.8 0.0925934016704559
1.8 0.158385321497917
};
\addplot [black, forget plot]
table {%
1.725 0.012775681912899
1.875 0.012775681912899
};
\addplot [black, forget plot]
table {%
1.725 0.158385321497917
1.875 0.158385321497917
};
\addplot [black, dotted, forget plot]
table {%
3.8 0.145129268643435
3.8 0.0203818663278583
};
\addplot [black, dotted, forget plot]
table {%
3.8 0.464982809656726
3.8 0.94379867162306
};
\addplot [black, forget plot]
table {%
3.725 0.0203818663278583
3.875 0.0203818663278583
};
\addplot [black, forget plot]
table {%
3.725 0.94379867162306
3.875 0.94379867162306
};
\addplot [black, dotted, forget plot]
table {%
5.8 0.140820821617185
5.8 0.021182405609226
};
\addplot [black, dotted, forget plot]
table {%
5.8 0.45481384274875
5.8 0.925775000592352
};
\addplot [black, forget plot]
table {%
5.725 0.021182405609226
5.875 0.021182405609226
};
\addplot [black, forget plot]
table {%
5.725 0.925775000592352
5.875 0.925775000592352
};
\addplot [black, dotted, forget plot]
table {%
0.2 0.0212724506855011
0.2 0.00569838285446167
};
\addplot [black, dotted, forget plot]
table {%
0.2 0.0425607413053513
0.2 0.0744854211807251
};
\addplot [black, forget plot]
table {%
0.125 0.00569838285446167
0.275 0.00569838285446167
};
\addplot [black, forget plot]
table {%
0.125 0.0744854211807251
0.275 0.0744854211807251
};
\addplot [black, dotted, forget plot]
table {%
2.2 0.0252844393253326
2.2 0.00604486465454102
};
\addplot [black, dotted, forget plot]
table {%
2.2 0.0484276115894318
2.2 0.0829941630363464
};
\addplot [black, forget plot]
table {%
2.125 0.00604486465454102
2.275 0.00604486465454102
};
\addplot [black, forget plot]
table {%
2.125 0.0829941630363464
2.275 0.0829941630363464
};
\addplot [black, dotted, forget plot]
table {%
4.2 0.0604170071171618
4.2 0.00573269093660078
};
\addplot [black, dotted, forget plot]
table {%
4.2 0.234566329952396
4.2 0.495606887941715
};
\addplot [black, forget plot]
table {%
4.125 0.00573269093660078
4.275 0.00573269093660078
};
\addplot [black, forget plot]
table {%
4.125 0.495606887941715
4.275 0.495606887941715
};
\addplot [black, dotted, forget plot]
table {%
6.2 0.0594548922064981
6.2 0.00554179909902575
};
\addplot [black, dotted, forget plot]
table {%
6.2 0.232155212442846
6.2 0.49113701076596
};
\addplot [black, forget plot]
table {%
6.125 0.00554179909902575
6.275 0.00554179909902575
};
\addplot [black, forget plot]
table {%
6.125 0.49113701076596
6.275 0.49113701076596
};
\path [draw=black, fill=babypink!60]
(axis cs:-0.35,0.0411806143820286)
--(axis cs:-0.05,0.0411806143820286)
--(axis cs:-0.05,0.0808447673916817)
--(axis cs:-0.35,0.0808447673916817)
--(axis cs:-0.35,0.0411806143820286)
--cycle;
\path [draw=black, fill=babypink!60]
(axis cs:1.65,0.0486404057592154)
--(axis cs:1.95,0.0486404057592154)
--(axis cs:1.95,0.0925934016704559)
--(axis cs:1.65,0.0925934016704559)
--(axis cs:1.65,0.0486404057592154)
--cycle;
\path [draw=black, fill=babypink!60]
(axis cs:3.65,0.145129268643435)
--(axis cs:3.95,0.145129268643435)
--(axis cs:3.95,0.464982809656726)
--(axis cs:3.65,0.464982809656726)
--(axis cs:3.65,0.145129268643435)
--cycle;
\path [draw=black, fill=babypink!60]
(axis cs:5.65,0.140820821617185)
--(axis cs:5.95,0.140820821617185)
--(axis cs:5.95,0.45481384274875)
--(axis cs:5.65,0.45481384274875)
--(axis cs:5.65,0.140820821617185)
--cycle;
\path [draw=black, fill=babyblue!60]
(axis cs:0.05,0.0212724506855011)
--(axis cs:0.35,0.0212724506855011)
--(axis cs:0.35,0.0425607413053513)
--(axis cs:0.05,0.0425607413053513)
--(axis cs:0.05,0.0212724506855011)
--cycle;
\path [draw=black, fill=babyblue!60]
(axis cs:2.05,0.0252844393253326)
--(axis cs:2.35,0.0252844393253326)
--(axis cs:2.35,0.0484276115894318)
--(axis cs:2.05,0.0484276115894318)
--(axis cs:2.05,0.0252844393253326)
--cycle;
\path [draw=black, fill=babyblue!60]
(axis cs:4.05,0.0604170071171618)
--(axis cs:4.35,0.0604170071171618)
--(axis cs:4.35,0.234566329952396)
--(axis cs:4.05,0.234566329952396)
--(axis cs:4.05,0.0604170071171618)
--cycle;
\path [draw=black, fill=babyblue!60]
(axis cs:6.05,0.0594548922064981)
--(axis cs:6.35,0.0594548922064981)
--(axis cs:6.35,0.232155212442846)
--(axis cs:6.05,0.232155212442846)
--(axis cs:6.05,0.0594548922064981)
--cycle;
\addplot [black, forget plot]
table {%
-0.35 0.0548651944845915
-0.05 0.0548651944845915
};
\addplot [black, forget plot]
table {%
1.65 0.0636655651032925
1.95 0.0636655651032925
};
\addplot [black, forget plot]
table {%
3.65 0.258364581615458
3.95 0.258364581615458
};
\addplot [black, forget plot]
table {%
5.65 0.24894283275682
5.95 0.24894283275682
};
\addplot [black, forget plot]
table {%
0.05 0.0286296010017395
0.35 0.0286296010017395
};
\addplot [black, forget plot]
table {%
2.05 0.0331724286079407
2.35 0.0331724286079407
};
\addplot [black, forget plot]
table {%
4.05 0.120636911907151
4.35 0.120636911907151
};
\addplot [black, forget plot]
table {%
6.05 0.118055251608637
6.35 0.118055251608637
};
\end{axis}

\end{tikzpicture}
    \caption{Performance metrics of AE vs. IDFT for $\text{SNR}=0$~dB.}
    \label{fig: box_plot}
    \vspace*{-4mm}
\end{figure}

\begin{figure}
    \centering
\begin{tikzpicture}[spy using outlines={rectangle, magnification=2, size=0.8cm, connect spies, }]
\definecolor{color0}{rgb}{0.843137254901961,0.0980392156862745,0.109803921568627}
\definecolor{color1}{rgb}{0.172549019607843,0.482352941176471,0.713725490196078}

\begin{axis}[
axis line style={white!15!black},
legend cell align={left},
legend style={nodes={scale=0.75, transform shape}, fill opacity=0.9, draw opacity=1, text opacity=1, at={(0.01,0.99)}, anchor=north west, draw=white},
tick align=outside,
tick pos=left,
x grid style={white!80!black},
xlabel={Average TX power [dB]},
grid = major,
grid style = {dotted, color = black, line width = 0.5pt},
xmin=-12, xmax=32,
xtick style={color=white!15!black},
xtick={-15,-10,-5,0,5,10,15,20,25,30,35},
xticklabels={\(\displaystyle -15\),\(\displaystyle -10\),\(\displaystyle -5\),\(\displaystyle 0\),\(\displaystyle 5\),\(\displaystyle 10\),\(\displaystyle 15\),\(\displaystyle 20\),\(\displaystyle 25\),\(\displaystyle 30\),\(\displaystyle 35\)},
y grid style={white!80!black},
ylabel={Per-user Rate [bpcu]},
ymin=-0.322990210067006, ymax=7.76019986809687,
ytick style={color=white!15!black},
ytick={0, 1, 2, 3, 4, 5, 6, 7},
yticklabels={\(\displaystyle 0\),\(\displaystyle 1\),\(\displaystyle 2\),\(\displaystyle 3\),\(\displaystyle 4\),\(\displaystyle 5\),\(\displaystyle 6\),\(\displaystyle 7\)},
]
\addplot [line width =1pt, color0]
table {%
-10 0.0512375247634074
-5 0.13415986490384
0 0.319906061481726
5 0.704004076241574
10 1.42220603975875
15 2.5601456191306
20 4.00443827915319
25 5.58927877107164
30 7.22528081553312
};
\addlegendentry{Perfect CSI, $\Delta f = 120$~MHz}
\addplot [line width =1pt, color0, dashed]
table {%
-10 0.0464864546132069
-5 0.122672236487982
0 0.29411604771588
5 0.650039970725216
10 1.31708189326173
15 2.37899710898227
20 3.71197994023056
25 5.08316101079865
30 6.29359143062981
};
\addlegendentry{AE, $\Delta f = 120$~MHz}
\addplot [line width =0.6pt, color0, dotted, mark=*, mark size=1.0, mark options={solid,fill opacity=0}]
table {%
-10 0.04638316854885
-5 0.122411903940674
0 0.293541319202687
5 0.648608901266808
10 1.31287758184259
15 2.36635452475915
20 3.67553512734716
25 4.98882251642702
30 6.08897051985005
};
\addlegendentry{AE, 8 bits, $\Delta f = 120$~MHz}
\addplot [line width =0.6pt, color0, dotted, mark=square, mark size=1.0, mark options={solid,fill opacity=0}]
table {%
-10 0.0460718704560854
-5 0.121641982886748
0 0.291666315270392
5 0.643799126059353
10 1.29894905025608
15 2.32575385805045
20 3.56596436486362
25 4.73403848302155
30 5.60936077719024
};
\addlegendentry{AE, 7 bits, $\Delta f = 120$~MHz}
\addplot [line width =1pt, color0, dotted]
table {%
-10 0.0444275207586243
-5 0.117009793641557
0 0.277595901321532
5 0.59915372972845
10 1.15485672721947
15 1.92386513660373
20 2.71258792555671
25 3.33412258766196
30 3.74163834067897
};
\addlegendentry{IDFT, $\Delta f = 120$~MHz}
\addplot [line width =1pt, color1]
table {%
-10 0.0566815775385364
-5 0.146576907538984
0 0.347404598977262
5 0.7583089024838
10 1.51781700591805
15 2.69566152167743
20 4.16104749644407
25 5.7540245565629
30 7.39278213727124
};
\addlegendentry{Perfect CSI, $\Delta f = 480$~MHz}
\addplot [line width =1pt, color1, dashed]
table {%
-10 0.0515961546544246
-5 0.134334343084418
0 0.31982422394764
5 0.700200337566222
10 1.40526001135231
15 2.50592555005716
20 3.85558973254944
25 5.22086361210997
30 6.40666175214709
};
\addlegendentry{AE, $\Delta f = 480$~MHz}
\addplot [line width =0.6pt, color1, dotted, mark=*, mark size=1.0, mark options={solid,fill opacity=0}]
table {%
-10 0.0515096117836581
-5 0.134103493062386
0 0.319284709778751
5 0.698750008170753
10 1.40123143616914
15 2.49426528964245
20 3.82253212504704
25 5.1360727458797
30 6.22432953766088
};
\addlegendentry{AE, 8 bits, $\Delta f = 480$~MHz}
\addplot [line width =0.6pt, color1, dotted, mark=square, mark size=1.0, mark options={solid,fill opacity=0}]
table {%
-10 0.0511966070287636
-5 0.133297100464975
0 0.317375355316711
5 0.693783686641062
10 1.38719216479334
15 2.45459401598857
20 3.71696859110268
25 4.89157299100104
30 5.76483398042876
};
\addlegendentry{AE, 7 bits, $\Delta f = 480$~MHz}
\addplot [line width =1pt, color1, dotted]
table {%
-10 0.049175065865896
-5 0.12859153236963
0 0.303492196639091
5 0.649634107563051
10 1.2450283363868
15 2.06094513483842
20 2.89037896381568
25 3.54489169606836
30 3.97683895698059
};
\addlegendentry{IDFT, $\Delta f = 480$~MHz}
\end{axis}
\spy[size=1.8cm,magnification=2.5] on (5.7,3.7) in node[fill=white] at (5.93, 0.91);
\end{tikzpicture}
    \caption{Per-user rate performance with LISA of DL CSI for \vari{$\mathbb{E}[\norm{\vH}^2_{\mathrm{F}}/ \norm{\vN}^2_{\mathrm{F}}] = 10$~dB} and a multi-user scenario with 8 users.}
    \label{fig: lisa_plot}
        \vspace*{-4mm}
\end{figure}
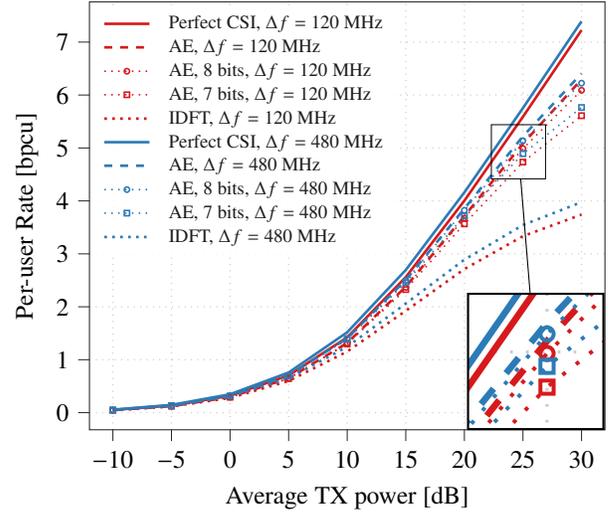

\section{Simulations}
\label{sec: simulations}
The autoencoder neural network has been implemented with Tensorflow~\cite{tensorflow2015} 
and single-precision has been utilized for the training. We consider mini-batches of $64$ samples and we use the Adam
optimization algorithm~\cite{adam} to tune the hyperparameters $\boldsymbol{\theta}$ and $\boldsymbol{\phi}$ of the neural network.
The weights are updated in order to minimize \wout{an empirical risk function based on the least-squares} loss function
\begin{equation}
\mathcal{L}(\boldsymbol{\theta}, \boldsymbol{\phi}) = \left\|\vg_{\boldsymbol{\phi}} \left(\vf_{\boldsymbol{\theta}} \left(\ulnoisy^{\text{real}}\right)\right) - \ulnoisy^{\text{real}}\right\|^2.
    \label{eq: loss_f}
\end{equation}
The UL-trained encoder is then used \wo{at each MT} to generate the codeword $\vz_{\text{DL}}$ from the noisy DL CSI estimate $\dlnoisy$. The codeword is then sent to the BS, which uses the UL-trained decoder to obtain a clean version of the DL CSI $\dlbighat \approxeq \dlbig$.

After the training, we measure the quality of the unsupervised denoising in terms of normalized mean square error $ \varepsilon^2 $ and cosine similarity $\rho$,
where
\begin{equation}
  \varepsilon^2 = \mathbb{E}\left[\frac{\norm{\hat{\vect{H}} - \vect{H}}_{\mathrm{F}}^2}{\norm{\vect{H}}_{\mathrm{F}}^2}\right]
\end{equation}
and
\begin{equation}
\rho = \mathbb{E}\left[\frac{1}{N_{\text{c}}}\sum_{n=1}^{N_{\text{c}}}\frac{\vert\hat{\vect{h}}^{\text{H}}_n\vect{h}_n\vert}{\norm{\hat{\vect{h}}_n}_2 \norm{\vect{h}_n}_2}\right],
\end{equation}
being $\vect{H} \in \mathbb{C}^{N_{\text{a}}\times N_{\text{c}}}$ the true CSI, and ${\vect{h}}_n$ its $n$-th column, and $\hat{\vect{H}}$ and
$\hat{\vect{h}}_n$ their corresponding versions at the decoder output. 

\wout{In addition,} we also evaluate the performance in terms of average per-user rate with zero forcing precoding. \wout{To this end,} we consider two different values of SNR, namely $10$~dB and $0$~dB, \vari{where the SNR represents the level of CSI corruption, \wout{i.e.,}  $\mathbb{E}[\norm{\vH}^2_{\mathrm{F}}/ \norm{\vN}^2_{\mathrm{F}}]$}.
\wout{We further} compare the results achieved with the UL-trained autoencoder with two methods that serve as \wout{a} reference. In particular, we utilize the CsiNet \wout{method}, which requires a learning-phase and has been proposed in~\cite{csinet}, and another method which is based on the IDFT which does not require any learning. \wout{CsiNet is based on an autoencoder approach trained on DL CSI that exploits the sparsity of CSI in the space-delay domain, and is often used as a benchmark.}
After transforming the DL CSI in the space-delay domain, the authors in~\cite{csinet} propose to retain only a small fraction of the component in the time domain, being the remaining component close to zero, and to train an autoencoder with this ``cropped'' version of the CSI. Specifically, we keep $64$ out of $160$ time-delay instances, and to be consistent with the original paper, only for the CsiNet results, we decide not to add any noise to the DL CSI.

For the approach based on the IDFT, first we transform the noisy DL CSI $\dlnoisy$ to the space-delay domain by a multiplication with a DFT matrix. Then, we only keep the first two columns in the space-time domain, such that the total number of coefficients is $256$, as it is assumed for the codeword. Afterwards, these coefficients are sent to the BS, which reconstructs the DL CSI in the space-frequency domain, by operating the zero-padding followed by the DFT transformation.
The results of NMSE and cosine similarity for $\text{SNR}= 10$~dB are displayed in \wout{the subplots of Fig.}~\ref{fig: cdfs}. We can clearly observe that the UL-trained autoencoder \wout{(``AE DL $120$ MHz'', ``AE DL $480$ MHz'')} performs very well on DL data too, with only a slight drop in performance when increasing the frequency gap \wout{from $120$ MHz to $480$ MHz.} {\wout{The ``AE UL'' curve demonstrates the reconstruction property of the autoencoder when applied to UL data, which serves as a further reference.}} Note that the \wo{other ``AE''-labeled solutions} have never seen training samples of DL CSI.  \wout{Nevertheless, it can be observed} that the ``AE'' solutions show \wout{considerable} gain compared to the \wout{``IDFT''} method and \wout{still some} gain compared to the \wout{``CsiNet'' curve}.
Analogous conclusions can be made by observing the performance metrics in Fig.~\ref{fig: box_plot} for $\text{SNR}= 0$~dB where the NMSE and cosine similarity achieved with our approach are compared with those of the IDFT approach.

Finally, \wout{results} of the average per-user rate in a multi-user scenario with $8$ users \wout{are discussed}. Likewise~\cite{utschick2021learning}, we adopt the LISA algorithm \cite{UtStJoLu18}
which is applied independently on
each of the $160$ carriers, and the results are then averaged over
the carriers.
Fig.~\ref{fig: lisa_plot} shows the per-user rate for $120$ and $480$ MHz frequency gaps, averaged over $100$ instances of LISA simulation runs for \vari{$\mathbb{E}[\norm{\vH}^2_{\mathrm{F}}/ \norm{\vN}^2_{\mathrm{F}}] = 10$~dB. 
} The continuous lines represent the rates
achievable with perfect DL CSI knowledge, the dashed lines represent the rates obtained with the
DL CSI predicted with \wo{the same UL-trained autoencoder at each MT}, and the dotted lines represent the rates with the IDFT method. We can observe that the rates per-user with the DL channels denoised with the AE is extremely close to the rates achieved with the true DL CSI and that there is a significant gain compared to the IDFT method.
Furthermore, we only notice a moderate degradation in the per-user rate when we apply uniform $8$ bit ($7$-bit) quantization to each element of the codewords, so that the total number of bits to be sent over the return channel is $256 \times 8 = 2048 $ bits ($256 \times 7 = 1792 $ bits). Note that the quantization of the codewords can be easily performed because the activation function at the end of the encoder forces the codeword values into the interval $[-1, 1]$.

\section{Conclusions}
\wout{In this work, following the idea of using autoencoders for noise reduction and codeword generation for DL CSI in FDD systems, we presented a novel concept. This is based on the recently discovered equivalence of UL and DL data across the FDD frequency gap, which allows training the autoencoder at the BS instead of the MT, followed by offloading the \wo{same} encoder to \wo{each} MT. Training on the MT is no longer necessary, making  it  possible  to  quickly  update the  encoder on the MT at any time and place. The promising results presented validate our proposed method.}

\bibliographystyle{IEEEtran}
\bibliography{IEEEabrv,new_ref}

\end{document}